\documentclass[twoside,twocolumn,10pt]{article}
\usepackage{graphicx}
\usepackage{amsmath}
\usepackage[colorlinks=true,linkcolor=blue,citecolor=blue]{hyperref}
\usepackage[margin=1in]{geometry}
\usepackage{setspace}
\usepackage{bm}
\usepackage{comment}
\usepackage[backend=biber,style=numeric,citestyle=numeric-comp,sorting=none]{biblatex}
\usepackage{siunitx}
\usepackage{abstract}
\usepackage{authblk}

\DeclareSIUnit\Molar{\textsc{m}}
\DeclareSIUnit\molar{\textsc{m}}
\DeclareSIUnit\enzymeunit{\textsc{units}}

\title{Mechanics of heterogeneous fiber networks processed by an active fluid}

\author[a,b]{Kyu Hwan Choi*}
\author[c,d]{Sattvic Ray*}
\author[c]{Reef Sweeney}
\author[c,e]{Zvonimir Dogic}
\author[a]{Sho C. Takatori\S}

\affil[a]{Department of Chemical Engineering, Stanford University, Stanford, CA 94305, USA}
\affil[b]{Department of Chemical Engineering, University of California, Santa Barbara, Santa Barbara, CA 93106, USA}
\affil[c]{Department of Physics, University of California, Santa Barbara, Santa Barbara, CA 93106, USA}
\affil[d]{Current address: Center for Living Systems, University of Chicago, Chicago, IL 60637, USA}
\affil[e]{Interdisciplinary Program in Quantitative Biosciences, University of California, Santa Barbara, Santa Barbara, CA 93106, USA}

\date{}
\begin{document}

\twocolumn[
    \begin{@twocolumnfalse}
    \maketitle
\text{*These authors contributed equally to this work}

\text{\S~Email: stakatori@stanford.edu}
    \begin{abstract}
    Internally generated active stresses can drive soft materials into structures that cannot form under thermal self assembly alone. For example, a microtubule-kinesin active fluid can assemble and irreversibly restructure actin-fascin networks. Increasing the motor concentration of the active fluid results in networks with broad pore size distribution and heterogeneous bundle structure. We probe the mechanics of these actively processed fiber networks by applying local forces with optical tweezers and measuring the strain response around the probe. Displacement fields of actively-processed networks propagate over a longer range than compared to unprocessed networks. Furthermore, networks processed at higher motor concentration are stiffer on average, while also exhibiting more heterogeneous response. At large strains, both unprocessed and actively processed networks exhibit strain softening and plastic restructuring. These measurements demonstrate that tunable active fluids can alter both the structure and viscoelastic properties of materials.
    \end{abstract}
    \end{@twocolumnfalse}
    \vspace{0.6cm}
    ]

\section{Introduction}

Heterogeneity is a ubiquitous feature of self-assembled fiber networks and colloidal gels~\cite{Gardel2004, lieleg2009structural, fleissner2016microscale,Shih1990,dinsmore2002direct,PatrickRoyall2008}. Such materials exhibit complex mechanical behaviors that are governed by their structural organization across multiple length scales~\cite{Schiessel1995, Broedersz2011, kim2014microstructure, arevalo2015stress, Ronceray2016, moghimi2017colloidal, Bantawa2023}. Macrorheology measures the bulk response of materials to global strain or stress applied at its boundaries. However, macrorheology does not elucidate how the material responds to forces applied at different length scales; it requires efficient coupling of the network to the boundaries, and it is challenging to implement on fragile samples with anisotropic mechanics. Alternatively, one-point microrheology reveals local rheological properties using the dynamics of embedded particles~\cite{mason1995optical, Mason1997}. Microrheology can characterize a limited amount of material and complex environments such as the cellular interiors\cite{lau2003microrheology}. It works well for materials where probe particles are larger than the mesh size. However, for heterogeneous materials, local effects determine the microrheological response. Passive two-point microrheology overcomes this limitation by measuring correlated thermal fluctuations of spatially separated probe particles, which recover the far-field mechanical response of the surrounding medium while minimizing probe-specific effects~\cite{crocker2000two,levine2001two,chen2003rheological,koenderink2006high}. However, passive two-point microrheology is limited to materials that are soft enough such that the embedded probe particles undergo measurable thermal fluctuations.

Here, we study the mechanics of stiff, athermal networks whose structure and spatial heterogeneity is controlled by microtubule–based active fluids~\cite{berezney2026active}. Active stresses drive rearrangements beyond those accessible through thermal fluctuations, producing networks with broad pore-size distributions. The resulting hierarchical and heterogeneous material reflects the history of active forcing. We characterize the mesoscale elasticity of such networks using optical tweezers while measuring the displacement field around the probe particle. Active processing enhances long-range strain propagation while increasing the mean and the variance of the local network stiffness. In the nonlinear strain regime, unprocessed and activity-processed networks exhibited plasticity and strain softening. In addition to demonstrating that actively-processed networks have spatially heterogeneous viscoelastic response, our work establishes an approach for measuring mesoscale mechanical response of heterogeneous materials.

\section{Experimental methods}
\subsection{Sample composition}
\textit{Active fluid}: We prepared a composite of microtubule (MT)-based active fluid and an actin-fascin system (Fig.~\ref{fig:sample-prep}A)~\cite{berezney2022extensile,berezney2026active}. Preparation of MT-based active fluid was described previously~\cite{tayar2022assembling,chandrakar2022engineering}. The kinesin-1 derived construct K401-BCCP was expressed in \textit{E. Coli.}~\cite{young1995subunit} The active fluid consisted of fluorescent MTs (\SI{25}{\micro\molar}, with \SI{3}{\percent} of monomers labeled with Alexa Fluor 647), K401-streptavidin motor clusters (KSA, 25-\SI{250}{\nano\molar}), and a truncated version of the MT-bundling crosslinker PRC1 (\SI{0.4}{\micro\molar})~\cite{subramanian2010insights}. A regeneration system composed of caged ATP (\SI{2.0}{\milli\molar}), PK/LDH (${\sim}$\SI{22}{\enzymeunit/\milli\liter}), and PEP (\SI{27}{\milli\molar}) maintained  constant activity. An antioxidant mixture of glucose (\SI{2.82}{\milli\gram/\milli\liter}), glucose oxidase (\SI{0.22}{\milli\gram/\milli\liter}), glucose catalase (\SI{0.04}{\milli\gram/\milli\liter}), and DTT (\SI{5.6}{\milli\molar}) minimized photobleaching. Stock solutions were stored in M2B buffer (\SI{80}{\milli\molar} PIPES, \SI{2}{\milli\molar} MgCl$_2$, \SI{1}{\milli\molar} EGTA, pH 6.8 with KOH). Additional MgCl$_2$ was added to a final concentration of \SI{5.3}{\milli\molar}.

\textit{Actin network}: The elastic network consisted of actin filaments and the actin-bundling crosslinker fascin. Actin was purified from rabbit muscle acetone powder (Pel-Freez), incubated with the non-hydrolyzable ATP analog adenylyl-imidodiphosphate (AMP-PNP), and diluted to final concentrations of 2-4 \SI{}{\micro\molar}. Fascin was used at final concentrations of 2-4 \SI{}{\micro\molar}, while phalloidin-Alexa 488 (Thermo Fisher) was kept constant at \SI{670}{\nano\molar} to label the actin filaments.

\textit{Gelsolin-coated beads}: Microbeads used for optical tweezing were bound to the actin-fascin network using gelsolin, an actin severing and end-binding protein. Gelsolin-coated beads can effectively transmit forces to attached actin filaments~\cite{Suzuki1996, Ward2015}. Gelsolin was purified from rabbit plasma (Pel-Freez) using a weak anion exchange column and stored in \SI{-80}{\celsius} in phosphate buffer~\cite{Kurokawa1990}. Gelsolin was subsequently attached to carboxyl-coated polystyrene beads (\SI{5}{\micro\meter}, Bangs Laboratories Product No.~13232). The gelsolin was conjugated to the bead's carboxyl groups using EDC (1-Ethyl-3-(3-dimethylaminopropyl)carbodiimide) NHS (N-hydroxysuccinimide) chemistry~\cite{Ward2015}. The gelsolin beads were suspended in M2B buffer, stored at \SI{4}{\celsius} and used for a few months.

\subsection{Sample preparation}
The sample mixture with all components except KSA was mixed with gelsolin-coated beads and divided into five aliquots of \SI{8.0}{\micro\liter} each.  KSA solution (\SI{2.9}{\micro\liter}) was added to each aliquot, resulting in a dilution series of samples with KSA concentrations of 190, 140, 90, and \SI{50}{\nano\molar}. An inactive control sample with \SI{0}{\nano\molar} KSA was included. The five samples were loaded into chambers (acrylamide-coated coverslips and parafilm spacers~\cite{lau2009condensation}) of \SI{2}{\milli\meter} width, \SI{24}{\milli\meter} length, and  \SI{120}{\micro\meter} height. The chambers were sealed with 5-minute epoxy and left for 1 hour at room temperature to allow for actin polymerization, fascin-induced bundling, and binding of gelsolin-coated beads to the actin-fascin network. During the incubation, the sample was placed on a rotator to prevent sedimentation of the beads and inactive MT-PRC1 bundles. The sample was then mounted on the microscope and exposed to UV light (8-12 \SI{}{\milli\watt/\milli\meter\squared}) for one minute, uncaging the ATP and initiating the active microtubule-kinesin flows.

\begin{figure}[ht!]
    \centering
    \includegraphics[width=\columnwidth]{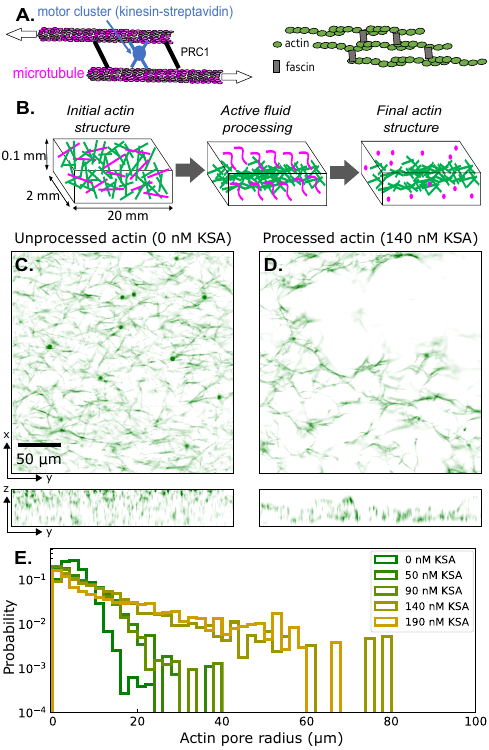}
    \caption{Microtubule-based active fluid restructures actin-fascin networks. (A) System components. Left: active fluid comprised of MTs, kinesin-streptavidin motor clusters, and the MT-bundling protein PRC1. Right: actin-fascin bundles. (B) Organization of MT (pink) and actin-fascin (green) bundles before, during, and after active processing. After processing, MTs are depolymerized, leaving behind actin-fascin structure in viscous solution. (C) Confocal z-slice (top) and x-slice (bottom) of network with no active processing. (D) Confocal slices of network after processing with active fluid. \SI{3.0}{\micro\molar} actin, \SI{1.5}{\micro\molar} fascin. (E) Pore size distribution of networks processed with varying amounts of KSA motor clusters. Probability is the area fraction of the 2D confocal slices that consist of detected 2D pores of the given radius.
    }
    \label{fig:sample-prep}
\end{figure}

After 24 hours, the sample was stored at \SI{4}{\celsius} overnight to halt any remaining activity. At this temperature, MTs depolymerized and bundles fell apart, while the actin-fascin network was left intact (Fig.~\ref{fig:sample-prep}B). Pore size distribution was computed from confocal z-scans of the static actin-fascin network (see SI for details). Optical tweezer measurements were performed at room temperature on the network within a few days. We did not observe a significant recovery of the MT-PRC1 bundles during this time.

\subsection{Optical tweezer measurements}

The gelsolin beads attached to the actin network were trapped using an optical tweezer (Tweez305, Aresis, Slovenia) with the \SI{5}{\watt} infrared ($\lambda$=\SI{1064}{\nano\meter}) laser. Objective lenses (CFI Apo $\lambda$S 40XC WI, N.A. 1.25; and CFI Plan Apochromat $\lambda$D 100X Oil, N.A. 1.45, Nikon) were used for strain and local viscoelastic response measurement, respectively.

\textit{Displacement field measurements}: The displacement fields induced by oscillating beads were measured at 3 beads each in the unprocessed (\SI{0}{\nano\molar} KSA) and processed (\SI{190}{\nano\molar} KSA) samples. Each \SI{5}{\micro\meter} bead was subject to an oscillating optical trap of \SI{5}{\micro\meter} amplitude at \SI{1}{\hertz}. Simultaneously, fluorescence images of the actin were acquired at 100 fps (Kinetix, Teledyne). The actin displacement field was calculated using the NCorr toolbox~\cite{blaber2015ncorr} (see SI for details).

\textit{Linear stiffness measurements:} The local network response was measured with optical tweezers, adapting active microrheology methods~\cite{bausch1999measurement,lee2010passive,chapman2014nonlinear}. Since it was not feasible to identify and simultaneously control two probe particles within the same optical plane at the desired separation, we used a single probe. We used gelsolin-coated beads that were specifically bound to actin filaments in the network, and thus transferred forces onto the network when pulled with the optical tweezers. This measurement is distinct from conventional active microrheology, which typically assumes a passivated probe that is much larger than the mesh size of the surrounding material. To measure the linear elastic response, the piezo stage (NanoScan SPZ600, Prior Scientific) was oscillated at $x_{\text{imposed}}(t) = x_0 \sin(\omega t)$, with $\omega=2\pi$~\si{\per\second} and $x_0$=\SI{0.5}{\micro\meter} while the optical trap was stationary. This defines a nominal imposed strain of $\gamma_0 = x_0/2R = 0.1$, based on the stage displacement. The particle motion was recorded at 100~fps (Prime95B, Teledyne).

We measured the optical trap stiffness to be  $\kappa_{\text{t}}=\SI{163.1}{}\pm\SI{1.6}{\pico\newton/\micro\meter}$ (Fig.~S5; see SI for trap calibration protocol). The force exerted on the probe particle was computed as $F_{\text{trap}}=-\kappa_{\text{t}}\Delta x$, where $\Delta x$ is the displacement of the particle center from the trap center. We report the diameter-normalized force $q(t)=F(t)/(2R)$, where $F(t)=F_{\text{trap}}(t)$ under the stated force-sign convention.
Because the probe is coupled to the surrounding network, its motion does not exactly follow the imposed stage displacement. We define the local strain of the network around the bead as $\gamma(t)=x_{\text{local}}(t)/(2R)$, where $x_{\text{local}}(t)=x_{\text{imposed}}(t)-\Delta x(t)$. We define $x_{\text{local,max}}=\max_t|x_{\text{local}}(t)|$ and $\gamma_{\text{max}}=x_{\text{local,max}}/(2R)$ for the centered oscillatory trajectory.
Using the imposed oscillation as the phase reference, the first-harmonic force is written as $F_1(t)=F_{\mathrm{st}}\sin(\omega t)+F_{\mathrm{lo}}\cos(\omega t)$, with $q_{\mathrm{st}}=F_{\mathrm{st}}/(2R)$ and $q_{\mathrm{lo}}=F_{\mathrm{lo}}/(2R)$. We define the local storage ($K'$) and loss ($K''$) stiffnesses as
\begin{equation}
K'=\frac{q_{\mathrm{st}}}{\gamma_{\text{max}}}
=\frac{F_{\mathrm{st}}}{x_{\text{local,max}}},\quad
K''=\frac{q_{\mathrm{lo}}}{\gamma_{\text{max}}}
=\frac{F_{\mathrm{lo}}}{x_{\text{local,max}}}.
\end{equation}

Frequency sweeps were performed at 0.1, 0.3, 1, 3, 10, and \SI{20}{\hertz} using 50 probe beads in each of the unprocessed and actively processed conditions at $x_0=\SI{0.5}{\micro\meter}$. Both $K'$ and $K''$ depended on frequency, and the two stiffness components crossed at approximately \SI{10}{\hertz} (Fig.~S11). Thus, a frequency of 1 Hz was chosen for most stiffness measurements because it was well below the crossover frequency and is in a regime where both stiffness components were measurable. Unless otherwise stated, the storage and loss stiffnesses reported in the main text were measured at \SI{1}{\hertz}.

\textit{Nonlinear stiffness measurements:} The optical trap was also used to measure nonlinear response of the actin network, following earlier approaches~\cite{lee2010passive, gupta2021optical}. The oscillation amplitude of the optical trap was set to \SI{5}{\micro\meter}, corresponding to a strain of $\gamma_0=1$. The imposed \SI{5}{\micro\meter} amplitude, defined as $x_0$, refers to the stage displacement, not the bead--trap offset $\Delta x$. The nonlinear response was analyzed only when $\Delta x$ remained within the displacement range ($\leq~\sim$\SI{0.5}{\micro\meter}) directly tested by the drag calibration. Lissajous curves were generated by plotting the normalized force $q=F(t)/(2R)$ against the imposed strain $x_{\mathrm{imposed}}(t)/(2R)$ during the oscillation cycles. To account for drift across cycles, Lissajous cycles were reconstructed using their first 20 Fourier modes (Fig.~S6,~S7). The tangential stiffness was computed as $K_{\mathrm{tan,imp}}=dF/dx_{\mathrm{imposed}}$, computed in the region $0.9x_0\leq|x_{\mathrm{imposed}}|\leq x_0$ ~\cite{wilhelm2002fourier,ewoldt2008new,hyun2011review,hyun2018nonlinear} (Fig.~S8). Note that this is distinct from the local differential stiffness, $dF/dx_{\mathrm{local}}$.

\section{Results}

\subsection{Active fluid drives structural remodeling of actin–fascin networks}
To generate networks of varying heterogeneity, we mixed a suspension of passive-actin-fascin bundles with an MT-kinesin active fluid (Fig.~\ref{fig:sample-prep}A)~\cite{berezney2026active}. Monomeric AMP-PNP actin was mixed with MT-based active fluid in the absence of hydrolyzable ATP. Actin polymerized around the passive MT-based active fluid, generating a homogeneous composite (Fig.~\ref{fig:sample-prep}B, left). In the absence of activity, both components exhibited no thermal motion. Photo-activation of the ATP induced kinesin-driven dynamics of MT bundles that continuously extended, buckled, frayed, and reformed, leading to spontaneous flows in the surrounding solvent and associated active stresses of 1-\SI{10}{\milli\pascal}~\cite{Gagnon2020, Adkins2022}. Autonomous flows rearranged actin filaments and fascin crosslinkers from their initial state into a multiscale heterogeneous network (Fig.~\ref{fig:sample-prep}B, center)~\cite{berezney2026active}. After consuming all the chemical fuel, active dynamics ceased after $\sim$12~hours. Subsequently, we disassembled MTs by cooling the sample to \SI{4}{\celsius} for 16~hours, leaving behind the static actin-fascin network (Fig.~\ref{fig:sample-prep}B, right). Without active processing, actin-fascin bundles formed a spatially heterogeneous fiber network (Fig.~\ref{fig:sample-prep}C). Active processing coarsened actin-fascin bundles while contracting the network towards the sample midplane (Fig.~\ref{fig:sample-prep}D). Activity yielded more heterogeneous networks with a broader pore size distribution (Fig.~\ref{fig:sample-prep}E).

\subsection{Actively processed networks exhibit enhanced strain propagation}

\begin{figure*}[ht!]
    \centering
    \includegraphics[width=\textwidth]{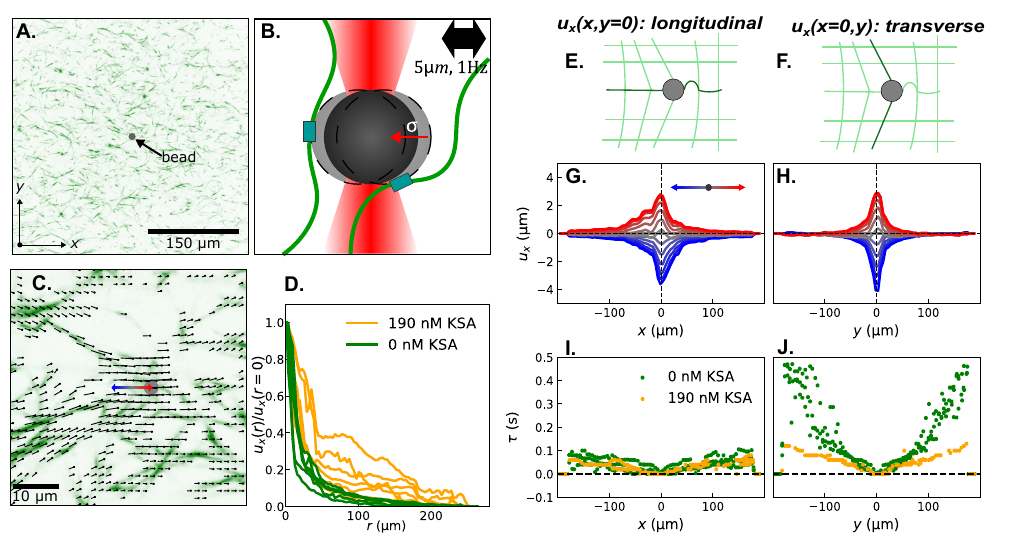}
    \caption{Active processing enhances strain propagation. (A) Actin network (z-slice) with a superimposed schematic of a bead (not to scale). (B) Schematic of bead in an optical trap bound to the network oscillating along the $x$-axis. The bead is attached with gelsolin (teal). (C) Displacement field (black arrows) of actin bundles (green) as the bead (gray) is moved to the right. (D) Actin displacements along the x-direction, $u_x$, as a function of distance from the bead center, $r$, after azimuthal averaging over the field of view. For each sample (\SI{0}{\nano\molar} KSA, \SI{190}{\nano\molar} KSA), the displacement fields around three separate beads were measured. The left-moving and right-moving periods of each oscillation cycle for a single bead are plotted as separate curves. (E) Hypothesized longitudinal deformations of the actin network (in black) subject to a right-moving bead. (F) Hypothesized transverse deformations (in black) due to the same bead motion. (G) Actin displacements $u_x$ in the x-direction as a function of longitudinal distance from the bead center $x$. For each value $x$, the field $u_x(x,y)$ was averaged near the x-axis ($|y|<\SI{32.5}{\micro\meter}$). Color indicates bead position during its oscillation period: red ($+x$) and blue ($-x$). Measurement was performed on a single bead in an unprocessed network. (H) The $u_x$ profile averaged along the x-axis for the same bead as in (G). (I) Temporal phase lag $\tau$ between the bead trajectory $x_\text{bead}(t)$ and the displacements $u_x(x,y,t)$ in the longitudinal direction. Measurements come from the same three beads in each sample as in (D). (J) Phase lag in the transverse direction. Measurements come from the same beads as in (I).}
    \label{fig:displacement-field}
\end{figure*}

We probed the elastic deformations of unprocessed and actively-processed networks with optical tweezers (Fig.~\ref{fig:displacement-field}A). The 10-\SI{100}{\micro\meter} average pore size precluded the use of traditional microrheology methods using passivated beads. Instead, we used gelsolin-coated beads bound specifically to the actin bundles (Methods). When a single gelsolin bead was oscillated with the optical trap, it pulled the attached actin bundles along with it and strained the surrounding actin network (Video~S2). We oscillated the \SI{5}{\micro\meter} beads by sinusoidally displacing the optical trap with an amplitude of \SI{5}{\micro\meter}. The resulting displacement field of the actin network around the bead resembled that of an elastic monopole~\cite{Landau1986} (Fig.~\ref{fig:displacement-field}B, C).

To investigate strain propagation in the networks, we oscillated the bead along the $x$-axis and measured the $u_x$ component of the displacement field as function of distance from the bead (Fig.~\ref{fig:displacement-field}D, Videos~S1,S3). This measurement demonstrated that bead-induced elastic deformations were more long-ranged in the actively processed networks. We fit the displacement field profiles to an inverse power law $u(r){\sim}r^{-p}$ (Fig.~S1). The unprocessed samples exhibited a power law of $p=1.2{\pm}0.1$, indicating more rapid spatial decay than in a continuum linear elastic medium (for which $p=1$). In contrast, the processed sample exhibited a displacement power law of $p=0.8{\pm}0.1$. These deviations from continuum linear elastic theory, in opposite directions, may reflect the mechanics of heterogeneous fiber networks at length scales where the continuum approximation breaks down~\cite{lerner2023anomalous,Goren2023}. We speculate that the sparse, thick bundles of the processed sample served as nearly-rigid cables that readily transmitted force, thus leading to enhanced strain propagation.

To explore the spatial anisotropy of the deformations, we compared their longitudinal and transverse modes (Fig.~\ref{fig:displacement-field}E,F). When a single probe particle in the unprocessed sample was displaced in the $+x$ direction, it induced a shorter-ranged $u_x$ deformation to its right and a longer-ranged deformation to its left (Fig.~\ref{fig:displacement-field}G, red lines). When the particle was displaced in the $-x$ direction, the shorter-ranged deformations were now to its left and the long-ranged deformations to its right (Fig.~\ref{fig:displacement-field}G, blue lines), suggesting that the actin network responds differently to tension and compression. The transverse displacement field of the same particle did not exhibit a noticeable asymmetry (Fig.~\ref{fig:displacement-field}H).

The displacement field propagated through the network with a temporal delay. At each point in the actin network, we defined $\tau(x,y)$ as the phase lag between the bead trajectory $x_\text{bead}(t)$ and the actin displacement field $u_x(x,y,t)$ (Fig.~S2). In both the unprocessed and processed samples, the phase lag $\tau$ increased with distance from the oscillating bead (Fig.~\ref{fig:displacement-field}I,J). Lags of similar value appeared to be localized on thick actin bundles, indicating rigid-like motion of these structures (Fig.~S3). We defined the longitudinal and transverse strain propagation speeds as the inverse slopes of the measured $\tau(x)$ and $\tau(y)$ curves, respectively (Fig.~S4). In the unprocessed sample, longitudinal strains propagated at \SI{2400}{}$\pm$\SI{600}{\micro\meter/\second} while transverse strains propagated at \SI{510}{}$\pm$\SI{60}{\micro\meter/\second}. In the actively processed sample, longitudinal strains propagated at \SI{2600}{}$\pm$\SI{300}{\micro\meter/\second}, while transverse strains propagated at \SI{1400}{}$\pm$\SI{100}{\micro\meter/\second}. Thus, the propagation speeds were slightly less anisotropic in the actively processed fiber network, perhaps a result of the structural changes induced by activity.

\begin{figure*}[ht]
    \centering
    \includegraphics[width=\textwidth]{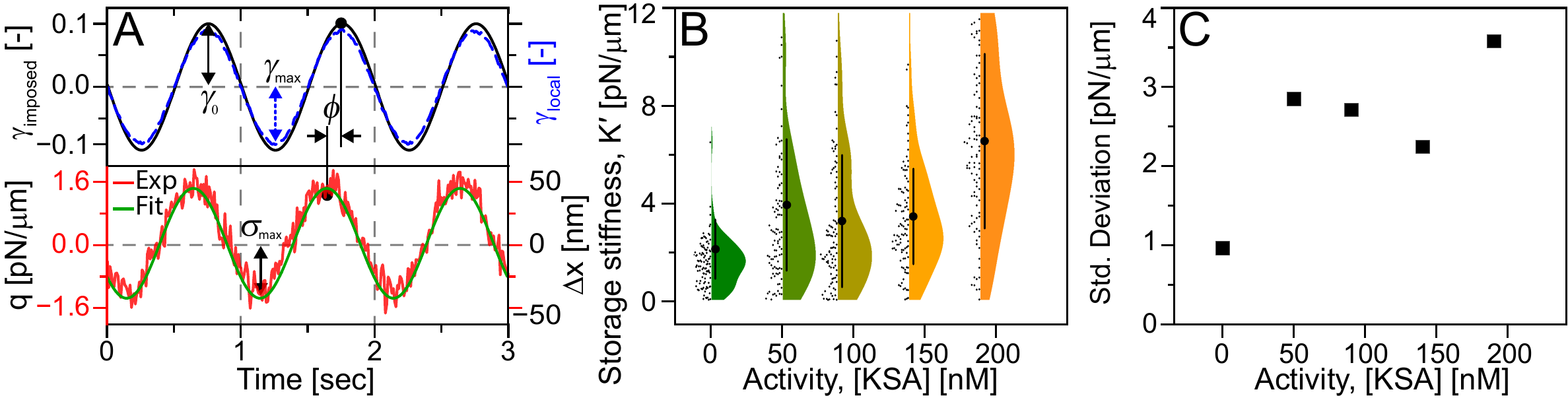}
    \caption{Active processing increases local network stiffness.
    (A) (Top) Imposed small-amplitude nominal strain ($\gamma_\text{imposed}$, black), $\gamma(t)=\gamma_0\sin{(\omega t)}$ with $\gamma_0=0.1$, and measured local strain ($\gamma_\text{local}$, dashed blue) obtained from the bead displacement. (Bottom) Normalized force response ( $q(t)=F(t)/(2R)$, red) on the trapped bead measured from the displacement ($\Delta x$) from the optical trapping center during oscillation. The normalized force is fitted to a sinusoidal function, $q(t)=q_{\text{max}}\sin(\omega t + \phi)$ (green), where $\phi$ is the phase lag between strain and normalized force, and $q_{\text{max}}$ is the maximum stress amplitude.
    (B) Distributions of local stiffness, $K'$, of the actin networks as a function of KSA motor concentration in the active fluid, extracted from force--strain measurements at individual beads. Each point is a measurement of network stiffness around a single bead. Approximately 100 beads were measured within a single network preparation at each KSA concentration. Error bars indicate standard deviations. Colored curves indicate kernel density estimates. (C) Standard deviations of the stiffness ($K'$) distributions as a function of KSA motor concentration.}
   \label{fig:micro-linear}
\end{figure*}

\subsection{Active processing locally increases network stiffness}
We compared the stiffnesses of the unprocessed and actively-processed actin networks. In analogy to bulk viscoelastic moduli, we defined local stiffness components $K'$ (storage) and $K''$ (loss) using the measured force and displacement on the beads as they were oscillated with the optical trap (Methods, Fig.~\ref{fig:micro-linear}A). We prepared an actin network at each of five KSA concentrations ranging from \SI{0}{\nano\molar} to \SI{190}{\nano\molar}. For each network, the local stiffness was sampled at $\sim$100 different beads (Fig.~\ref{fig:micro-linear}B). The unprocessed network had a mean storage stiffness of $\sim$\SI{2.0}{\pico\newton/\micro\meter}, whereas the network processed at the highest activity (\SI{190}{\nano\molar} KSA) had a mean storage stiffness of $\sim$\SI{7.1}{\pico\newton/\micro\meter}. The standard deviations of the stiffness distributions generally increased with KSA concentration (Fig.~\ref{fig:micro-linear}C). The local loss stiffness $K''$ showed a similar mean increase, from $\sim$\SI{1.2}{\pico\newton/\micro\meter} in the unprocessed network to $\sim$\SI{3.1}{\pico\newton/\micro\meter} at \SI{190}{\nano\molar} KSA, with concomitant increase in standard deviation (Fig.~S10). These measurements demonstrate that increased active processing results in actin networks that are both stiffer and more heterogeneous in the spatial distribution of stiffness. Taken together with the activity-induced pore size heterogeneity (Fig.~\ref{fig:sample-prep}E), the stiffness measurements suggest that structural and mechanical heterogeneities in the
fiber networks are closely linked, and that active stresses generate locally stiff regions with thick, load-bearing bundles embedded within softer surroundings.

\subsection{Actin-fascin networks plastically deform and strain-soften}
\begin{figure}[ht!]
    \centering
    \includegraphics[width=\columnwidth]{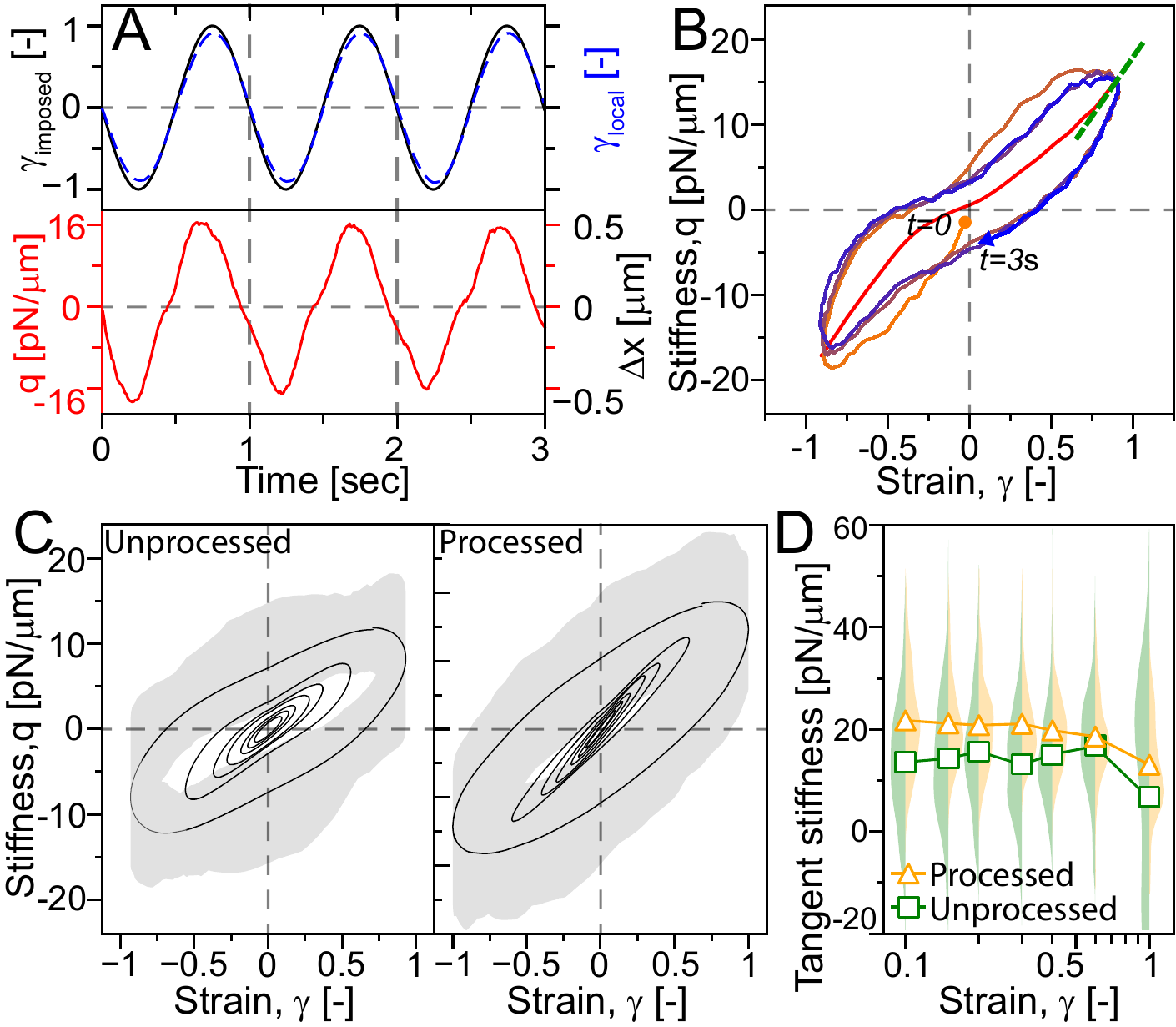}
    \caption{Non-linear mechanics, plasticity, and strain softening of heterogeneous networks.
    (A) Imposed strain (black, $\gamma_\text{imposed}$) and measured local strain (dashed blue, $\gamma_\text{local}$) on the bead at $\gamma_0=1$, and corresponding normalized-force response and bead displacement ($\Delta x$) (red) showing multi-mode wobbles characteristic of nonlinear response.
    (B) Lissajous curves of diameter-normalized force $q=F/(2R)$ versus nominal imposed strain $\gamma_{\mathrm{imposed}}=x_\text{imposed}/(2R)$ from panel (A). The colored (orange to blue) line represents a large strain cycle $\gamma_0=1$, showing hysteresis.
    The red line is the mid-line extracted from the Fourier fitting. The green dashed line is the imposed-displacement tangential stiffness $K_{\mathrm{tan,imp}}=dq/d\gamma_{\mathrm{imposed}}$ at the loop edge ($|\gamma|=0.9\gamma_\text{max}-\gamma_\text{max}$). (C) Lissajous curves averaged across multiple beads in the unprocessed (left) and actively-processed \SI{190}{\nano\molar} KSA (right) networks. For each network, 50 beads were sampled. Shaded regions represent the 20–80th percentile envelope of curves from individual bead trajectories.
    (D) Strain-dependent tangential stiffness $K_{\mathrm{tan,imp}}$ of the unprocessed (yellow) and actively-processed (green) networks. Shaded envelopes show the full distribution of extracted imposed-displacement tangential stiffnesses at each strain amplitude. The distributions broaden with strain as different regions stiffen or soften.}
    \label{fig:micro-nonlinear}
\end{figure}

To measure the nonlinear response of the actin networks, we increased the sinusoidal oscillation amplitude of the beads to \SI{5}{\micro\meter} ($\gamma_0 = 1$). In this regime, the force on the bead exhibited multiple harmonics (Fig.~\ref{fig:micro-nonlinear}A). The Lissajous (force-displacement) curve of a single bead changed in shape over multiple oscillation periods, indicating plastic restructuring (Fig.~\ref{fig:micro-nonlinear}B). The nonzero area inside the Lissajous curve indicated hysteresis, possibly due to viscous dissipation or plastic rearrangement of the network. To compare the nonlinear response of unprocessed and processed networks, we measured Lissajous curves at multiple strain amplitudes ($\gamma_0:0.1-1$) for 50 beads in each sample (Fig.~\ref{fig:micro-nonlinear}C). To account for drift across multiple cycles of each Lissajous curve, we extracted only the periodic component of the force signal (Fig.~S6, S7). Beads probed in both samples exhibited a wide distribution of nonlinear force-extension curves (Fig.~S8,~S9). This heterogeneity suggested that the local actin structure around each individual bead greatly affected its response (Videos~S4,~S5). Furthermore, we measured the tangential stiffness $K_{\mathrm{tan,imp}}=dq/d\gamma_{\mathrm{imposed}}$ from the median line slope of the reconstructed Lissajous curve near the maximum imposed strain amplitude ($|\gamma_{\mathrm{imposed}}|\sim\gamma_0$) (Fig.~\ref{fig:micro-nonlinear}B and Fig.~S8). On average, both unprocessed and actively-processed networks exhibited modest strain softening at $\gamma_0\approx1$ (Fig.~\ref{fig:micro-nonlinear}D). However, both networks also exhibited a broad distribution in local stiffness, with some beads exhibiting strain stiffening and others strain softening (Fig.~S8). Thus, active processing had little effect on the nonlinear mechanics of the fiber networks.

\section{Conclusion}
We used optical tweezers to probe the mechanics of actin-fascin fiber networks processed by the microscopic stresses generated by an MT-kinesin active fluid. At higher kinesin concentration, the active fluid sculpted increasingly heterogeneous fiber networks. When subject to a point force, the unprocessed network exhibited a displacement field with power law $u(r)\sim 1/r^{1.2}$, while the actively-processed network exhibited a slower decay of $u(r)\sim 1/r^{0.8}$. Previous work found that the displacement decayed as $u(r)\sim1/r^{0.6}$ along the perpendicular axis of uniaxially stretched fibrin hydrogels~\cite{Goren2023}. Such anomalous long-range elasticity has been theoretically derived for fiber networks comprised of linear elements but near a critical rigidity transition~\cite{lerner2023anomalous}. We speculate that the actively processed actin networks in our study may fall in a similar regime.

In the unprocessed network, longitudinal displacement modes propagated faster than transverse modes, while in the actively processed network, both modes propagated at similar speed. It is unclear what sets the propagation speeds in either network, but they may be related to the theoretically predicted bias towards compression in fiber networks~\cite{Ronceray2016}, or to the finding that fibrin hydrogels respond more strongly in the direction perpendicular to applied tension~\cite{Goren2023}.

By measuring the force response of several beads attached to the fiber networks, we found that MT fluids with higher active stresses generated stiffer actin-fascin networks. However, it must be noted that our measurements of network stiffness are local and sampled primarily in actin-rich regions where the beads are located. Thus, they may not represent the macroscopic stiffness of the entire network. In particular, the large porous regions, which are not sampled in our measurements, may actually result in a decreased global stiffness of the networks formed at higher activity.

Our measurements suggest that the important mechanical characterization for heterogeneous fiber networks is not the mean stiffness, but rather the distribution of stiffness values across the network. Further measurements will be required to determine if stiffness distribution correlates more with the KSA concentration or with statistics of the network structural heterogeneity. It would also be instructive to determine whether beads embedded in actin-rich or actin-sparse regions of the network are consistently stiffer or softer. It is possible that different regions are closer to the transition between linear and nonlinear response due to fiber organization and prestress in the network.
In particular, the strain-softening behavior observed at an imposed oscillation amplitude of  \SI{5}{\micro\meter} (Fig.~\ref{fig:micro-nonlinear}) suggests that nonlinear mechanical effects may contribute to the displacement fields measured at the same amplitude (Fig.~\ref{fig:displacement-field}). However, the precise onset of nonlinear response cannot be determined from these measurements alone. These higher-order couplings between structure and mechanics require future investigation.

The activity-induced network stiffening suggests that active processing involves a self-limiting mechanical feedback mechanism~\cite{berezney2026active}. Chaotic active flows cause initially small actin-fascin bundles to bind together, resulting in a network of large bundles interspersed with empty regions. This plastic restructuring continues until the network is locally stiff enough to resist further deformation by active flows, reminiscent of the stiffening of bundled actin networks due to external cyclic shear~\cite{schmoller2010cyclic}. Theoretical work on the effect of active stresses on fiber networks has also predicted activity-induced network stiffening~\cite{Broedersz_MacKintosh_2011, Ronceray2016, Kumar2026}. However, these and related works are generally inspired by actin-myosin networks and thus model active forces as \textit{contractile} dipoles embedded in the network. In contrast, the activity in our experimental system is generated by \textit{extensile} MT-kinesin bundles. Further theoretical work is required to elucidate the interactions between active extensile and passive adhesive bundles.

This work complements a recent study of actin--microtubule composites activated by KSA motor clusters~\cite{Sheung2025}. In that study, MT and actin filaments increasingly demixed as the KSA concentration was increased, resulting in greater structural heterogeneity. Network stiffness was measured with passivated beads in the actively restructuring composite, with maximal stiffness reported at intermediate motor concentrations. In contrast, the system studied here consists of actin and MT \textit{bundles} rather than individual filaments, and we measure the mechanics of the static actin network after active MTs are effectively removed from the system. Further work is necessary to establish a comprehensive picture of the structure, dynamics, and mechanics accessible with cytoskeletal composites.

Active matter processing opens up the possibility of exquisitely tailored fiber networks. Theoretical studies on idealized models of fiber networks predict rich mechanical properties~\cite{Broedersz2011, Zhou2018, Bantawa2023}. So far, experimental realizations of such mechanics are limited due to the difficulty of constructing macroscopic networks with controlled microstructure~\cite{SoareseSilva2011, Alvarado2013, Sharma2016}. Our results demonstrate that active stresses can rearrange microscopic fiber networks at the scale of individual fibers and pores, tuning both structure and mechanical response. To achieve greater control over actively generated network structures, spatiotemporal modulation of active flows may be implemented using, for example, light-sensitive motors and methods of control theory~\cite{ross2019controlling, Norton2020, Lemma2023, Nishiyama2025}. Our results are a first step towards harnessing active stresses to assemble fiber networks with tunable mechanical properties.

\section*{Author contributions}
All authors conceptualized the work;
K.H.C., S.R. and R.S. performed experiments and analyzed the data;
Z.D. and S.C.T. supervised the study;
All authors wrote the paper.

\section*{Data availability}
Data for this article, including processed microscopy data, structural characterization data, local probe response measurements, nonlinear force--displacement analysis data, and analysis code, are available at Zenodo at: \href{https://doi.org/10.5281/zenodo.20128984}{10.5281/zenodo.20128984}. The Supplementary Information and supporting videos are provided as ancillary files with this submission.

\section*{Conflicts of interest}
There are no conflicts to declare.

\section*{Acknowledgments}
This work was primarily supported by the National Science Foundation through the Materials Research Science and Engineering Center (MRSEC) at UC Santa Barbara: NSF DMR–2308708 (IRG-2). The preparation of elastic networks and the characterization of strain propagation was supported by the Department of Energy Office of Basic Energy Sciences through project DE-SC0022291.

\printbibliography

\end{document}